\newcommand{\be}{\begin{equation}}
\newcommand{\ee}{\end{equation}}
\newcommand{\bea}{\begin{eqnarray}}
\newcommand{\eea}{\end{eqnarray}}
\newcommand{\beas}{\begin{eqnarray*}}
\newcommand{\eeas}{\end{eqnarray*}}
\newcommand{\nn}{\nonumber}

\documentclass[onecolumn,aps,showpacs,nofootinbib,epsfig,axodraw]{revtex4}
\usepackage{graphics}
\usepackage{epsfig}
\usepackage{axodraw}
\begin{document}
\title{3-Point Off-shell Vertex in Scalar QED in Arbitrary Gauge and Dimension}
\author{A. Bashir$^\dagger$, Y. Concha-S\'anchez$^{\dagger}$ and 
     R. Delbourgo$^\ddagger$}
\affiliation{$^\dagger$Instituto de F{\'\i}sica y Matem\'aticas,
Universidad Michoacana de San Nicol\'as de Hidalgo, Apartado Postal
2-82, Morelia, Michoac\'an 58040, M\'exico.\\
$^\ddagger$ School of Mathematics and Physics, University of Tasmania, Locked 
Bag 37 GPO, Hobart 7001, Australia.}

\begin{abstract}
We calculate the complete one-loop off-shell three-point scalar-photon vertex 
in arbitrary gauge and dimension for Scalar Quantum Electrodynamics. 
Explicit results are presented for the particular cases of dimensions 3 and 4 
both for massive and massless scalars. We then propose non-perturbative forms
of this vertex that coincide with the perturbative answer to order $e^2$.
\end{abstract}


\maketitle

\section{Introduction}

The non-perturbative structure of Green functions in gauge field theories has 
turned out to be a challenging problem. Aside from the complicated non-abelian
scenario of Quantum Chromodynamics (QCD), even simpler examples such as 
Quantum Electrodynamics (QED) have proved a hard nut to crack in the 
non-perturbative regime. Nevertheless, gauge covariance relations, such as the 
Ward-Fradkin-Green-Takahashi Identity (WFGTI)~\cite{WFGT}, and the 
Landau-Khalatnikov-Fradkin transformations (LKFT)~\cite{LKF} contain vital 
clues about the Green functions. Guided by such relations, extensive work has 
been carried out to construct non-perturbative Green functions
\cite{GT,WFGTIappl, LKFappl}. As well, perturbation theory is a reliable 
guide when constraining acceptable structures in the weak coupling 
limit~\cite{BC, KRP, BKP0, BKP1, BKP2, BR1}.

In the context of perturbation theory, a systematic study of spinor QED was 
initiated by Ball and  Chiu~\cite{BC}. They decomposed the vertex 
into a `longitudinal' part which ensures that the WFGTI identity is satisfied, 
and a `transverse' part. In a basis where kinematic singularities are avoided, 
they gave off-shell results for the one-loop transverse vertex in 4 dimensions 
in the Fermi-Feynman gauge. Later on, K{\i}z{\i}lers\"{u}, Reenders and 
Pennington extended this result to an arbitrary covariant gauge~\cite{KRP}. 
Results for massless and massive QED3 were obtained afterwards
\cite{BKP1,BKP2,BR1,ALV}. These results were then generalized to arbitrary 
dimension by  Davydychev, Osland and Saks~\cite{davydychev} in the 
realm of QCD (from which all QED results can be inferred). Whereas the bare 
fermion-boson vertex in a minimal coupling gauge theory is merely $\gamma^{\mu}$,
in general the vertex can be expanded out in terms of 12 spin amplitudes 
constructed from $\gamma^{\mu}$ and two independent four-momenta~\cite{Bernstein}.
The WFGTI fixes four coefficients of the twelve spin amplitudes in terms of the 
fermion functions comprising the longitudinal component. The transverse part thus
involves eight vectors with eight unconstrained scalar coefficients that 
depend on the gauge parameter $\xi$, the space-time dimension $d = 2\ell$, 
fermion masses and three kinematical invariants ($k^2,p^2,q^2$); so this is a 
complicated problem even at one-loop order.

One might hope that, in the absence of spinorial matrices, Scalar Quantum 
Electrodynamics (SQED) can offer a simpler platform to study non-perturbative 
solutions~\cite{salam}. In this scenario, the 3-point vertex can be written in 
terms of just two independent four-momenta. The WFGTI fixes the coefficient of 
one of these. Therefore, there is only one unconstrained function which defines 
the transverse vertex---representing an 8-fold simplification of spinor QED/QCD! 
The trade-off is that additional four-point interactions occur in SQED. Thus the 
1-loop scalar-photon vertex involves two additional Feynman diagrams. Ball and 
Chiu~\cite{BC} carried out this calculation for massive scalars in the 
Fermi-Feynman gauge ($\xi = 1$) for $d = 4$. In this article, we extend their work 
to arbitrary dimension $d$ and gauge $\xi$ involving the one-loop scalar 
propagator along the way.

There are several reasons why this calculation is helpful: (i) it keeps track of 
the correct gauge covariance properties of the Green functions; (ii) one can take 
on-shell limits to check the gauge invariance of physical observables---this 
would not be not possible~\cite{davydychev} if one only has results near
four dimensions; (iii) SQED anyway has interest in lower dimensions, for example
non-perturbative SQED in 2+1- and 0+1-dimensions has been examined by
\cite{Waites} and~\cite{SGT} respectively; (iv) three dimensional field theories 
contain several features of corresponding four dimensional field theories at 
high temperatures~\cite{Gross}.

We have organized the article as follows. In Section II we introduce the 
notation to calculate the three-point vertex, discuss its decomposition in the 
light of WFGTI and give the expressions for one-loop scalar propagator and the 
longitudinal component of the three-point vertex. In Section III we evaluate the
complete one-loop vertex in arbitrary gauge and dimensions and hence deduce an 
expression for its transverse component. We suggest three simple and natural 
constructions for its non-perturbative counterpart in Section IV and finish by 
discussing the so-called transverse Takahashi identities in Section V.
An appendix serves to summarise many useful expressions arising from the
Feynman integrals.

\section{Preliminaries}
We shall start by setting out the notation, discussing the WFGTI and decomposing 
the 3-point vertex into longitudinal and transverse components. We then make use 
of the scalar propagator to present the longitudinal part fully at 1-loop order. 
Constant reference is made here and in the next section to various (mainly 
hypergeometric) functions which are listed in an appendix.

\subsection{Notation}
   We define the bare quantities in the usual form:  the scalar propagator
$S^0(p)=  1/(p^2-m^2)$, the photon propagator
$\Delta^0_{\mu \nu} = -\left[ g_{\mu \nu} p^2-(1-\xi) p_{\mu}p_{\nu}\right]/p^4$,
the 3-point vertex $\Gamma^0_{\mu} = (k+p)_{\mu}$ and the 4-point double
photon vertex $e^2 \Gamma^0_{\mu \nu}= e^2 g_{\mu \nu}$, where $\xi$
is the general covariant gauge parameter (such that $\xi=0$ corresponds
to Landau gauge) and $e$ is the usual QED coupling constant. The 3-point
vertex up to one-loop order is diagrammatically represented in Figure 1;
it can be written in terms of two vectors alone, namely, $k^{\mu}$ and $p^{\mu}$
or, if preferred, $P^\mu \equiv (p+k)^\mu$ and $q^\mu \equiv (k-p)^\mu$.
Due to the presence of the four-point vertex, there are two additional diagrams 
to be calculated in addition to the usual one required for the spinor QED.

\vspace{0.5cm}
\begin{center}
\begin{picture}(2000,70)(13,-30)
\SetScale{0.78}
\SetWidth{1.2}
\CCirc(90,15){5}{Black}{Black}
%\Vertex(226,15){2} \Text(226,25)(0)[]{$\mu$}
%\Vertex(346,15){2} \Text(346,25)[]{$\mu$}
%\Vertex(256,-104){2} \Text(266,-104)[]{$\alpha$}
%\Vertex(371,40){2}  \Text(381,40)[]{$\alpha$}
%\Vertex(371,-10){2} \Text(381,-10)[]{$\beta$}
%\Vertex(226,-135){2} \Text(226,-125)[]{$\mu \nu$}
%\Vertex(346,-135){2} \Text(346,-125)[]{$\mu \nu$}
%\Vertex(376,-165){2} \Text(386,-166)[]{$\alpha$}
\SetColor{Red}
\Photon(25,15)(85,15){4}{7}
\Photon(166,15)(226,15){4}{7}
\SetColor{Black}
\LongArrow(65,25)(45,25) \PText(55,5)(0)[]{q=k - p}
\LongArrow(206,25)(186,25) \PText(196,5)(0)[]{q=k - p}
\LongArrow(326,25)(306,25) \PText(324,5)(0)[]{q=k - p}
\LongArrow(386,5)(386,25) \PText(396,15)(0)[]{w}
\LongArrow(350,25)(360,35)
\LongArrow(360,-5)(350,5)
\SetColor{Blue}
\ArrowLine(93,20)(143,65)
\ArrowLine(143,-35)(93,10)
\ArrowLine(226,15)(276,65)
\ArrowLine(276,-35)(226,15)
\SetColor{Black}
\PText(146,75)(0)[]{p}
\PText(146,-45)(0)[]{k}
\PText(146,15)(0)[]{ = }
\PText(276,75)(0)[]{p}
\PText(276,-45)(0)[]{k}
\PText(276,15)(0)[]{+}
\SetColor{Red}
\Photon(296,15)(346,15){4}{7}
\SetColor{Blue}
\ArrowLine(346,15)(396,65)
\ArrowLine(396,-35)(346,15)
\SetColor{Black}
\PText(396,75)(0)[]{p}
\PText(396,-45)(0)[]{k}
\SetColor{Red}
%controversial photon
\Photon(376,45)(376,-15){3}{6.5}
\SetColor{Black}
\PText(416,15)(0)[]{+}
\PText(350,40)(0)[]{p-w}
\PText(350,-10)(0)[]{k - w}
%Last two diagrams
\LongArrow(466,25)(446,25)
\LongArrow(586,25)(566,25)
\LongArrow(495,32)(505,42)
\LongArrow(620,-10)(610,0)
\LongArrowArc(503,30)(10,257,-15)
\LongArrowArc(626,0)(10,-13,95)
\PText(464,5)(0)[]{q=k - p}
\PText(584,5)(0)[]{q=k - p}
\SetColor{Red}
\Photon(426,15)(486,15){4}{7}
\SetColor{Blue}
\ArrowLine(486,15)(536,65)
\ArrowLine(536,-35)(486,15)
\SetColor{Black}
\PText(536,75)(0)[]{p}
\PText(536,-45)(0)[]{k}
\SetColor{Red}
\PhotonArc(500,30)(20,225,45){2}{6.5}
\SetColor{Black}
\PText(525,15)(0)[]{w} \PText(546,15)(0)[]{+}
\PText(500,50)(0)[]{p-w}
\SetColor{Red}
\Photon(556,15)(606,15){3}{7}
\SetColor{Blue}
\ArrowLine(606,15)(656,65)
\ArrowLine(656,-35)(606,15)
\SetColor{Black}
\PText(656,75)(0)[]{p}
\PText(656,-45)(0)[]{k}
\SetColor{Red}
\PhotonArc(623,0)(20,-45,135){2}{6.5}
\SetColor{Black}
\PText(648,15)(0)[]{w}
%\PText(726,15)(0)[]{+}
\PText(613,-15)(0)[]{k-w}
\SetScale{0.9}
\SetColor{Black}
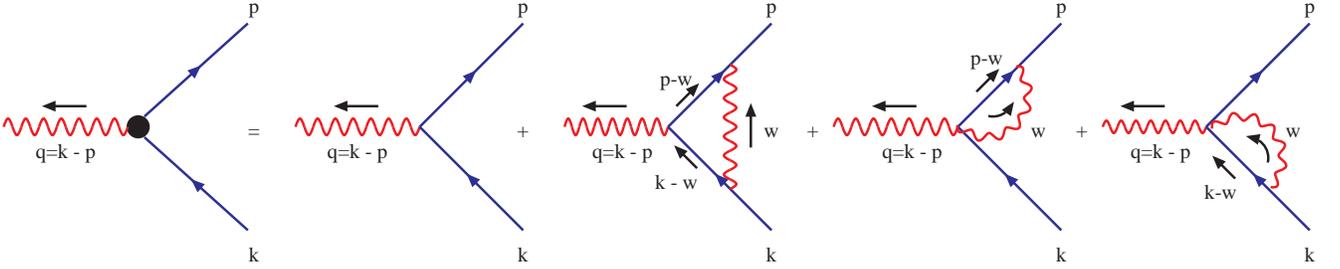
\PText(290,-70)(0)[]{Figure 1: One loop 3-point vertex in SQED.}
\end{picture}
\end{center}
\vspace{2.0cm}

The full 3-point vertex satisfies the usual WFGTI:
\begin{equation}
    q_{\mu} \Gamma^{\mu}(k,p) = S^{-1}(k) - S^{-1}(p)  \;
\end{equation}
and has the non-singular limit
\begin{equation}
      \Gamma^{\mu}(p,p) = \partial S^{-1}(p)/\partial p_\mu \;
\end{equation}
when $k \rightarrow p$.
We can use (1) to construct the `longitudinal' part of the vertex~:
\begin{equation}
      \Gamma^{\mu}_L(k,p) = \frac{S^{-1}(k)-S^{-1}(p)}{k^2-p^2}\;(k+p)^\mu\;.
\end{equation}
The full vertex can then be written as
\begin{equation}
        \Gamma^{\mu}(k,p) = \Gamma^{\mu}_L(k,p) + \Gamma^{\mu}_T(k,p) \;,
\end{equation}
where the `transverse' part satisfies
\begin{equation}
        q_{\mu} \Gamma^{\mu}_T(k,p)=0 \;,\qquad \qquad \Gamma^{\mu}_T(p,p)=0\;,
\end{equation}
and can be expanded out only in terms of the basis vector
\begin{eqnarray}
         T^{\mu}(k,p)= k\cdot qp^{\mu} - p\cdot q k^{\mu}=
         [q^\mu (k^2-p^2)-(k+p)^\mu q^2]/2 \;, \label{basis}
\end{eqnarray}
Thus the full vertex is
\bea
\Gamma^{\mu}(k,p) &=& \frac{S^{-1}(k) - S^{-1}(p)}{k^{2} - p^{2}}\;(p+k)^\mu 
  + \tau(k^{2},p^{2},q^{2}) \;  T^{\mu}(k,p) \;. \label{LT}
\eea
The coefficient $\tau$ is a Lorentz scalar function of $k$ and $p$, and can
be expressed in terms of the 3 invariants, $k^2, p^2$ and $q^2$. Thus knowing
only one unknown function $\tau$ is sufficient to fix the full 3-point vertex 
completely in SQED, the rest being tied to the scalar propagator $S$.

\subsection{Longitudinal Vertex}

At one-loop the scalar propagator is given by two diagrams but because massless 
tadpole type diagrams are zero in dimensional regularization (which we are
adopting), only the first 
diagram contributes.
\begin{center}
\begin{picture}(100,80)(0,0)
\SetWidth{1.2}
\Vertex(20,15){2}
\Vertex(80,15){2}
\Vertex(192,15){2}
\SetColor{Blue}
\ArrowLine(-150,15)(-100,15)
\ArrowLine(-100,15)(-50,15)
\CCirc(-100,15){5}{Black}{Black}
\ArrowLine(0,15)(100,15)
\Text(-25,15)[]{$\;$ = $\;$}
\Text(50,5)[]{$k$}
\SetColor{Red}
\PhotonArc(50,15)(30,0,180){4}{8.5}
\Text(120,15)[]{$\;$ + $\;$}
%\Text(55,25)[]{$iS^{0}$}
\SetColor{Black}
\LongArrowArc(50,15)(20,60,120)
\LongArrowArc(190,30)(20,60,120)
\SetColor{Blue}
\ArrowLine(140,15)(190,15)
\ArrowLine(190,15)(240,15)
\SetColor{Red}
\PhotonArc(190,40)(20,-90,90){4}{8}
\PhotonArc(190,40)(20,90,-90){4}{8}
\SetColor{Black}
%\LongArrow(30,17)(45,25)
%\LongArrow(232,10)(232,-10)
%\LongArrow(255,40)(275,40)
%\LongArrow(20,10)(20,-10)
%\LongArrow(80,10)(80,-10)
%\LongArrowArcn(85,20)(20,120,60)
\Text(50,55)[]{$k-p$}
%\Text(10,20)[]{$\mu$}
%\Text(90,20)[]{$\nu$}
\Text(-100,0)[]{$p$}
\Text(10,5)[]{$p$}
\Text(90,5)[]{$p$}
\Text(190,70)[]{$k$}
\Text(193,5)[]{$p$}
%\Text(232,-20)[]{$2ie^{2}g^{\mu \nu}$}
%\Text(293,40)[]{$i\Delta_{\mu \nu}^{0}$}
%\Text(80,-20)[]{$-ie\Gamma_{0}^{\nu}$}
%\Text(20,-20)[]{$-ie\Gamma_{0}^{\mu}$}
%\Text(110,38)[]{$i\Delta^{0}_{\mu \nu}$}
\SetScale{0.9}
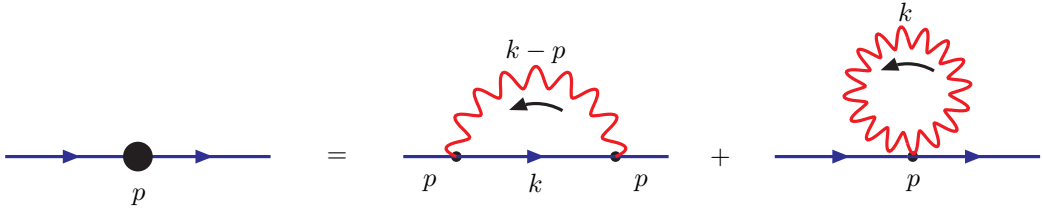
\PText(50,-40)(0)[]{Figure 2: One-loop scalar propagator.}
%\Text(50,-40)[c]{Figure 2: One-loop scalar propagator.}
\end{picture}
\end{center}
\vspace{2cm}

In arbitrary dimensions $d=2\ell$ and gauge $\xi$, the inverse propagator 
at one-loop is given by
\bea
\nn
S^{-1}  =  
% p^{2} - m^{2} 
 \frac{-e^{2}}{m^2}\left( \frac{m^2}{4 \pi} \right)^{\!\ell} 
 \Gamma \left(1 - \ell\right) \bigg\{ 1 - 2 \frac{(m^{2} + p^{2})}{m^2}
 {}_2F_{1}\bigg(2-\ell,1;\ell;\frac{p^{2}}{m^{2}}\bigg) 
 +(1\!-\!\xi)\frac{(m^2\!-\!p^2)^2}{m^4}{}_2F_{1}\bigg(3-\ell,2;\ell;
   \frac{p^{2}}{m^{2}}\bigg)\!\bigg\} 
\label{oneloopFP} \\ 
\eea
and readily yields the longitudinal part of the 3-point vertex at one-loop~:
\bea
\Gamma_{L}^{\mu}(k,p) &=&
\frac{e^{2}\pi^{2} (k+p)^\mu}{(2\pi)^{2\ell}(k^2\!-\!p^2)}
 \bigg\{\!2Q_{1}(k)(m^2\!+\! k^2)-\!2Q_{1}(p)(m^2\!+\!p^2)+(1-\xi)
[(m^2\!-\!p^2)^2Q_{3}(p)-(m^2\!-\!k^2)^{2}Q_{3}(k)]\!\bigg\},\label{LongVert}
\eea
where the functions $Q_i(p)$ are tabulated in the appendix.

\section{One-Loop Vertex}

In this section, we shall evaluate the complete off-shell one-loop vertex 
for all $\xi$ and $\ell$. Subtracting the longitudinal part from that
produces the remaining transverse part.

\subsection{The Full Vertex}

The complete one-loop correction to the vertex is the sum of the
three contributions that correspond to the last three graphs of Figure 2~:
\bea \Lambda^{\mu}=\Lambda_{1}^{\mu}(k,p)
+ \Lambda_{2}^{\mu}(p) + \Lambda_{2}^{\mu}(k)\;, \label{vertex}
\eea
The first contribution involving only 3-point vertices is given by~:

\bea
\Lambda^{1}_{\mu} \hspace{-1mm} &=& \hspace{-1mm}\frac{-ie^{2}}{(2\pi)^{2\ell}}
\bigg\{4(k\cdot p)(k + p)_{\mu}J^{(0)} + \left[ -8(k\cdot p)g_{\mu}^{\;\;\;\nu}
-2(k + p)_{\mu}(k + p)^{\nu} \right] J_{\nu}^{(1)}
+ 4(k + p)^{\nu}J_{\mu \nu}^{(2)} +(k + p)_{\mu}K^{(0)} -2K_{\mu}^{(1)} 
\nn \\ && \hspace{-1cm}
+(\xi - 1)\left[(k+p)_{\mu}K^{(0)}+4(k+p)_{\mu}p^{\alpha}k^{\beta}I_{\alpha
\beta}^{(2)}-8p^{\alpha}k^{\beta}I^{(3)}_{\mu\alpha \beta}-2(k+p)_\mu
(k+p)^{\alpha}J_{\alpha}^{(1)} 
%\nn \\&& \hspace{15mm}
+ 4(k + p)^{\alpha}J^{(2)}_{\mu \alpha}  - 2K^{(1)}_{\mu} \right] \bigg \}\;,
\eea
where
\bea
K^{(0)}&=&\int d^{d}w
\frac{1}{[(p - w)^{2} - m^{2}][(k - w)^{2} - m^{2}]} \nn \\
K_{\mu}^{(1)} &=& \int d^{d}w \frac{w_{\mu}}{[(p-w)^2-m^2][(k-w)^2- m^2]}\nn \\
 \nn \\ \nn
J^{(0)} &=& \int d^d w \frac{1}{w^{2}[(p-w)^2-m^2][(k-w)^2-m^2]} \\ \nn \\ \nn
J_{\mu}^{(1)} &=& \int d^{d}w \frac{w_{\mu}}{w^{2}[(p - w)^{2} - m^{2}][(k - w)^{2} - m^{2}]} \\ \nn \\ \nn
J_{\mu\nu}^{(2)}&=& \int d^{d}w \frac{w_{\mu}w_{\nu}}{w^{2}[(p - w)^{2} - m^{2}][(k - w)^{2} - m^{2}]}  \\ \nn \\ \nn
I^{(0)}&=&\int d^{d}w \frac{1}{w^{4}[(p-w)^2-m^2][(k-w)^2-m^2]}\\ \nn \\ \nn
I_{\mu}^{(1)}&=&\int d^{d}w\frac{w_{\mu}}{w^{4}[(p-w)^2-m^2][(k-w)^2-m^2]}
\\ \nn \\ \nn
I^{(2)}_{\alpha \beta}&=&\int d^{d}w
\frac{w_{\alpha}w_{\beta}}{w^{4}[(p - w)^{2} - m^{2}][(k - w)^{2} -
m^{2}]} \\ \nn \\
I_{\mu \alpha \beta}^{(3)}&=&\int d^{d}w
\frac{w_{\mu}w_{\alpha}w_{\beta}}{w^{4}[(p - w)^{2} - m^{2}][(k -
w)^{2} - m^{2}]} \;. \label{TheIntegrals}  \\ \nn
\eea
The results of computing these integrals are provided in detail in the appendix 
where we also compare them with other calculations in the literature for some 
particular cases of $d$. The two $\Lambda_2^{\mu}$ contributions
contain the 4-point vertices. They are relatively simple to evaluate as they 
contain only propagator type loops. Thus we only quote the final result~:
\bea
\Lambda_2^{\mu}(p)&=&\frac{e^{2}\pi^{2}p^\mu}{(2\pi)^{2\ell}}\bigg \{
\bigg[3+ \frac{m^{2}}{p^{2}} \bigg ]Q_{1}(p)-\frac{\pi^{\ell-2}}{p^{2}}\Gamma
(1 - \ell)(m^{2})^{\ell - 1}
+(\xi -1)\frac{(p^{2} - m^{2})}{p^{2}}[Q_{1}(p) + (p^{2}
-m^{2})Q_{3}(p)]\bigg\}. \label{23}
\eea
Eqs.~(\ref{vertex}--\ref{23}) form the complete one-loop scalar-photon vertex 
for any $\xi$ and $\ell$ at the one-loop level. This is a generalization to 
arbitrary dimension and gauge of the work of Ball and Chiu \cite{BC} who only 
examined the case $\xi=1$, $\ell = 2$, using cut-off regularization. The 
explicit answers for the integrals (12) are stated in the appendix; as a 
general abbreviation we will write $X^{(0)} =i\pi^2X_0/2$ in what follows and 
use $\{d,m\}$ as a superscript (or subscript) to signify dimension and mass, 
as and when needed.

\subsection{The Transverse Vertex}

The transverse vertex is obtained by subtracting the longitudinal vertex,
Eq.~(\ref{LongVert}), from the full one, Eqs.~(\ref{vertex}--\ref{23}), at 
one-loop. Carrying out this exercise, we arrive at the following coefficient 
$\tau$ of the transverse vector $T^{\mu}$ for massive scalars~:
\bea
\tau_{d,m}(k^2,p^2,q^2)
&=&\frac{e^{2}\pi^{2}}{2(2\pi)^{d}\Delta^{2}}\bigg\{(k^2 - 2m^2 + 
 p^2 - 4k\cdot p)\left[ -K_{0} + (m^2 + k\cdot p)J_{0}\right]\nn \\ && +
\frac{2Q_1(p)}{k^2-p^2} \left[ p^{2}(p^2-3k\cdot p) + k^2(k\cdot p-3p^2)-
 2m^2(p^{2} + k\cdot p) \right] \nn \\
&& -\frac{2Q_{1}(k)}{k^2 - p^2}\left[ k^2(k^2-3k\cdot p)+p^2(k\cdot p-3k^2) 
-2m^{2}(k^{2} + k\cdot p)\right] \nn \\
&& +(\xi -1)(m^{2} -k^{2})(m^{2} - p^{2}) \Big[J_{0} - (k\cdot p + m^{2})I_0 -
 \nn \\ 
&&\hspace{4.5cm}\frac{2Q_{3}(p)}{k^{2} - p^{2}}(k\cdot p + p^{2}) +
\frac{2Q_{3}(k)}{k^{2} - p^{2}}(k\cdot p + k^{2}) \Big] \bigg\} \; ,
\label{taudm} 
\eea 
where $I_0, J_0, K_0, Q_{1-6}$ are explicitly stated in the appendix. 
$\Delta^2 \equiv(k \cdot p)^2-k^2 p^2 = (k\cdot q)^2-k^2q^2$. Thus
$4\Delta^2=\lambda(k^2,p^2,q^2)=[p^4+k^4+q^4-2p^2k^2-2p^2q^2-2k^2q^2]$, 
the K\"all\'en function and is related to the (2$\times$area)$^2$ of a triangle 
with sides $\sqrt{k^2}, \sqrt{p^2}, \sqrt{q^2}$. In the massless limit, the above 
expression reduces to
\bea
\tau_{d,0}&=&\frac{e^{2}\pi^{2}}{2(2\pi)^{d}\Delta^{2}}\bigg\{(k^{2}
+ p^{2} -4k\cdot p)[(k\cdot p)J_{0}^{d,0} - K_{0}^{d,0}] 
\nn \\ && \hspace{1.8cm}
+\frac{2Q_{1}^{d,0}(p)}{k^{2} - p^{2}}[p^{4} -3(k^{2} + k\cdot
p)p^{2} + k^{2}k\cdot p]
%\nn \\ && 
-\frac{2Q_{1}^{d,0}(k)}{k^{2} -
p^{2}}[k^{4} -3(p^{2} + k\cdot p)k^{2} + p^{2}k\cdot p]
\nn \\ && \hspace{1.8cm}
+(\xi - 1)k^{2}p^{2}\bigg[J_{0}^{d,0} - k\cdot pI_{0}^{d,0} 
%\nn \\ && 
-\frac{2Q_{3}^{d,0}(p)}{k^{2} - p^{2}}(k\cdot p + p^{2}) +
\frac{2Q_{3}^{d,0}(k)}{k^{2} - p^{2}}(k\cdot p + k^{2})\bigg] \bigg
\}  \;.  \label{taud0} 
\eea 
In the massive case for small $\epsilon = 2-\ell$ one gets 
\bea \tau_{4 -2\epsilon,m}
&=&
\frac{\alpha}{8 \pi \Delta^{2}}\bigg\{(k^{2} - 2m^{2}
+ p^{2} - 4k\cdot p)[(m^{2} + k\cdot p)J_{0}^{4 - 2\epsilon,m}+ 2{\cal S}] 
\nn \\ && \hspace{1cm}
+ \frac{2L(p)}{k^{2} - p^{2}}\bigg(p^{2}(p^{2} - 3k\cdot p) +
k^{2}(k\cdot p - 3p^{2}) - 2m^{2}(p^{2} + k\cdot p)\bigg) 
\nn \\ && \hspace{1cm}
- \frac{2L(k)}{k^{2} - p^{2}}\bigg(k^{2}(k^{2} - 3k\cdot p) +
p^{2}(k\cdot p - 3k^{2}) - 2m^{2}(k^{2} + k\cdot p)\bigg)
\nn \\ && \hspace{1cm}
+ (\xi - 1)(m^{2} - k^{2})(m^{2} - p^{2})\bigg[J_{0}^{4 - 2\epsilon,m}
- \frac{2}{\chi}(k\cdot p + m^{2})\bigg(-q^{2}{\cal S} 
\nn \\ && \hspace{1cm}
+ \frac{p^{2}[(p^{2} - m^{2})q^{2} + 2m^{2}(k^{2} - p^{2})]L(p)}{(p^{2} -
m^{2})^{2}} + \frac{k^{2}[(k^{2} - m^{2})q^{2} - 2m^{2}(k^{2} -
p^{2})]L(k)}{(k^{2} - m^{2})^{2}} \bigg) 
\nn \\ && \hspace{1cm}
+ 2\frac{(k\cdot p + p^{2})(p^{2} + m^{2})}{(k^{2} - p^{2})(p^{2} - m^{2})^{2}}L(p)
-2\frac{(k\cdot p + k^{2})(k^{2} + m^{2})}{(k^{2} - p^{2})(k^{2} -
m^{2})^{2}}L(k) \bigg]\bigg\} \;.
\eea 
The quantities ${\cal S}$, $L$ can be found early on in the appendix. 
In the massless limit for small $\epsilon$ one reads off
\bea \tau_{4 - 2\epsilon,0}&=&\frac{\alpha}{8 \pi
\Delta^{2}}\bigg \{(k^{2} + p^{2} - 4k \cdot p)\bigg(k\cdot p
J_{0}^{4 - 2\epsilon,0} + \ln\frac{q^{4}}{p^{2}k^{2}}\bigg)
+ \frac{(k^{2} + p^{2})q^{2} - 8p^{2}k^{2}}{p^{2} - k^{2}}\ln
\frac{k^{2}}{p^{2}} 
\nn \\ && \hspace{1cm}
+ (\xi - 1)k^{2}p^{2}\bigg[J_0^{4 -2\epsilon,0}+ 
2\frac{p^{2}(k^{2} + k\cdot p)\ln(-p^{2})}{k^{2}- p^{2}} 
- 2\frac{k^{2}(p^{2} + k\cdot p)}{k^{2} -p^{2}}\ln(-k^{2}) + 
2k\cdot p\ln(-q^{2})\bigg]\bigg\}  \;.
\eea 
Note that for $\xi=1$, it agrees with Eq.~(2.9) of~\cite{BC}.
In the massive case with  $d=3$ we end up with
\bea 
\tau_{3,m}&=&\frac{e^{2}}{16\pi
\Delta^{2}}\bigg \{(k^{2} - 2 m^{2} + p^{2} - 4k\cdot p)
\left[(m^{2} + k \cdot p)J_{0}^{3,m} - 2I(\frac{q^{2}}{4}) \right]
\nn \\ && \hspace{0cm}
+ 4(k^{2}(k^{2} - 3k\cdot p) - 2(k^{2} + k\cdot p)m^{2} 
+ (k\cdot p - 3k^{2})p^{2})\frac{I(k^{2})}{(k^{2} -
p^{2})} 
\nn \\ && \hspace{0cm}
-4(k^{2}(k\cdot p - 3 p^{2}) + p^{2}(p^{2} - 3k\cdot p) 
- 2m^{2}(k\cdot p + p^{2}))\frac{I(p^{2})}{(k^{2}
- p^{2})} 
\nn \\ &&  \hspace{0cm}
+ (\xi - 1)(m^{2} - k^{2})(m^{2} - p^{2})\bigg [ J_{0}^{3,m} 
  - \frac{4m(k^{2} + k\cdot p)}{(k^{2} -
m^{2})^{2}(k^{2} - p^{2})} + \frac{4m(p^{2} + k\cdot p)}{(m^{2} -
p^{2})^{2}(k^{2} - p^{2})} 
\nn \\ && \hspace{0cm}
-\frac{1}{\chi}\bigg[2(m^{2} + k\cdot
p)
  \bigg(\frac{J_{0}}{2}(m^{2} +
k\cdot p)q^{2} + m\bigg(\frac{(k^{2} - m^{2})q^{2} - (k^{2} +
m^{2})(k^{2} - p^{2})}{(k^{2} - m^{2})^{2}} 
% \nn \\ && 
- \frac{k^{2} - p^{2} + q^{2}}{m^{2} - p^{2}}\bigg)\bigg)\bigg]
\bigg ]\bigg \} ,
\eea 
where $\chi$ is a geometrical quantity listed in the appendix.
The result (18) simplifies remarkably in the massless limit~: 
\bea
\tau_{3,0}&=&\frac{e^2}{2{\cal K}{\cal P}{\cal Q}} \left[
 \frac{{\cal K}^2 + 2{\cal Q}{\cal K} + {\cal P}^2 + 2{\cal P}{\cal Q}}
 {({\cal K}+{\cal P})({\cal K} + {\cal P} + {\cal Q})} + 
 (\xi - 1)\frac{1}{4}\right] \;,  \label{tau30} 
\eea
where we have adopted the Euclidian notation $\sqrt{-k^{2}}={\cal K}$,
$\sqrt{-p^{2}}={\cal P}$, $\sqrt{-q^{2}}={\cal Q}$.

\section{On the Non-perturbative Vertex}

The one-loop expression for $\tau(k^2,p^2,q^2)$ provides a guide as to its 
possible form in the strong coupling regime. Any non-perturbative {\em ansatz} 
for the transverse vertex should reduce to the perturbative result evaluated 
above. Eqs.~(\ref{oneloopFP}) and~(\ref{taudm}) suggest what $\tau$ might 
resemble for general $e^2$~:
\bea
\tau_{d,m}(k^2,p^2,q^2)
&=&\frac{1}{4\Delta^{2}} \; \frac{\left[ S^{-1}(k,\xi=1)-S^{-1}(p,\xi=1) \right]}{
\left[ (m^2+k^2)Q_1(k)-(m^2+p^2)Q_1(p) \right]} \; \times
\nn \\ && \hspace{1cm}
\bigg\{(k^{2} - 2m^{2} + p^{2} - 4k\cdot p)  
 \left[ -K_{0} + (m^{2} + k\cdot p)J_{0} \right] 
\nn \\ && \hspace{1cm}
+ \frac{2Q_{1}(p)}{k^{2} - p^{2}}
\left[ p^{2}(p^{2} - 3k\cdot p) + k^{2}(k\cdot p -3p^{2}) - 2m^{2}(p^{2} + k\cdot p) \right] 
\nn \\ && \hspace{1cm}
-\frac{2Q_{1}(k)}{k^{2} - p^{2}}
\left[ k^{2}(k^{2} - 3k\cdot p) + p^{2}(k\cdot p - 3k^{2}) -2m^{2}(k^{2} + k\cdot p)\right] \bigg\}
\nn \\&& \nn \\
&+&  \frac{1}{2\Delta^{2}} \; \frac{\left[ S^{-1}(k,\xi-1)-S^{-1}(p,\xi-1) \right]}{
\left[ (m^2-k^2)^2Q_3(k)-(m^2-p^2)^2Q_3(p) \right]} \;  (m^{2} -k^{2})(m^{2} - p^{2}) \; \times
\nn \\&& \hspace{1cm}
\bigg\{J_{0} - (k\cdot p + m^{2}) I_{0} 
- \frac{2Q_{3}(p)}{k^{2} - p^{2}}(k\cdot p + p^{2}) +
\frac{2Q_{3}(k)}{k^{2} - p^{2}}(k\cdot p + k^{2})\bigg\} \;.
\label{taudmnp} 
\eea 
The notation here means that $S(p,\xi=1)$ is the scalar propagator 
in the Fermi-Feynman gauge, whereas, $S(p,\xi-1)$ is the coefficient of the 
scalar propagator proportional to $(\xi-1)$. By construction, this expression 
reproduces the one-loop transverse vertex in the weak coupling regime. In 
specific dimensions and for the massless case, it simplifies. Of 
course its form is not unique but it is perhaps the simplest non-perturbative 
extension of our earlier results for any $\xi$ and $d$. We expect that an 
identical two-loop calculation will help us pin down the exact structure better. 
Due to the lack of Dirac matrices, this two-loop calculation is not as 
formidable a task as for spinor QED or QCD. We are currently in the process of 
carrying it out.

Another approach is to tie in the asymptotic behaviour with the anomalous
dimension of the scalar field in 4-d. To see how this is done, return to the 1-loop
self energy as obtained previously in eq. (8) ; by using contiguity relations
of hypergeometric functions, this can be cast in the simpler form
\bea
\Sigma(p) &=&  
 \frac{-e^2}{m^2}\left( \frac{m^2}{4 \pi} \right)^{\!\ell} 
 \Gamma(1 - \ell) \bigg[ 1+2(\ell-1)(1-\xi)+
 \{(1-2\ell)-\xi(3-2\ell)\}\left(1+\frac{p^2}{m^2}\right)
   {}_2F_1\bigg(2-\ell,1;\ell; \frac{p^{2}}{m^{2}}\bigg)\!\bigg] 
\label{Sigma}\nn
\eea
According to the procedure for self-consistent regularization by higher-order
corrections \cite{RD} we then make the substitution $\ell\rightarrow2+\gamma$ in 
eq. (\ref{Sigma}) and renormalize by ensuring that the propagator behaves 
as $(-p^2)^{1+\gamma}$ as $p^2\rightarrow\infty$. This gives the self-consistent
asymptotic equation,
\bea
\left(-\frac{p^2}{m^2}\right)^{1+\gamma}&\sim& \frac{e^2}{16\pi^2}
 \frac{\Gamma^2(1+\gamma)}{\Gamma(2\gamma+2)}\bigg[(\xi-3)\Gamma(-\gamma)
 +2(1-\xi)\Gamma(1-\gamma)\bigg]\left(1+\frac{p^2}{m^2}\right)
 \left(1-\frac{p^2}{m^2}\right)^\gamma,
\eea
which fixes $\gamma=(3-\xi)e^2/16\pi^2 +{\rm O}(e^4)$. But anyway it produces
a non-perturbative form of the 4-d propagator
\bea
S^{-1}(p) &\simeq & \frac{e^2}{16(\gamma+1)\pi^2}
    \bigg[(\xi-3)\Gamma(-\gamma) +2(1-\xi)\Gamma(1-\gamma)\bigg]
    \left(1+\frac{p^2}{m^2}\right){}_2F_1(-\gamma,1;\gamma+2;p^2/m^2).
\eea
[This non-perturbative method succeeds in the ultraviolet but fails in the 
infrared limit $p^2\rightarrow m^2$, when he propagator $S\sim 
1/(p^2-m^2)^{1+(3-\xi)e^2/8\pi^2}$.
For infrared exponentiation it is much  easier to resort to the gauge 
technique~\cite{GT}.]
Exactly the same procedure can be applied to the transverse vertex. If $\tau$
is expressed in Feynman parametric form, 
\bea
(p^2-k^2)\tau(p,k,q) &=& \frac{4e^2\Gamma(2-\ell)}{(4\pi)^\ell}\int_0^1 d\sigma\,
(1-\sigma)^{\ell-2}\bigg[(m^2-p^2\sigma)^{\ell-2}-(p\rightarrow k)\bigg] \nn\\
&&+ \frac{e^2(2m^2+k^2+p^2-2q^2)\Gamma(3-\ell)}{(4\pi)^\ell}\int_0^1
d\sigma\, (1-\sigma)^{\ell-1}\int_{-1}^1 du\, u{\cal D}^{\ell-3}\nn\\
&&+\frac{e^2(\xi-1)(k^2-m^2)(p^2-m^2)\Gamma(4-\ell)}{(4\pi)^\ell}\int_0^1
d\sigma\,(1-\sigma)^{\ell-2}\int_{-1}^1 du\, u{\cal D}^{\ell-4},
\eea
with ${\cal D} \equiv m^2-q^2(1-\sigma)(1-u^2)/4-p^2\sigma(1+u)/2-k^2\sigma(1-u)$,
then with the above substitution, an ansatz for the non-perturbative transverse 
vertex in 4-d emerges~:
\bea
(p^2-k^2)\tau(p,k,q) &=& \frac{4e^2\Gamma(-\gamma)}{(4\pi)^2}\int_0^1 d\sigma\,
(1-\sigma)^{\gamma}\bigg[(m^2-p^2\sigma)^{\gamma}-(p\rightarrow k)\bigg] \nn\\
&&+ \frac{e^2(2m^2+k^2+p^2-2q^2)\Gamma(1-\gamma)}{(4\pi)^2}\int_0^1
d\sigma\, (1-\sigma)^{1+\gamma}\int_{-1}^1 du\, u{\cal D}^{\gamma-1}\nn\\
&&+\frac{e^2(\xi-1)(k^2-m^2)(p^2-m^2)\Gamma(2+\gamma)}{(4\pi)^2}\int_0^1
d\sigma\,(1-\sigma)^\gamma\int_{-1}^1 du\, u{\cal D}^{\gamma-2},  \label{Anom-vert}
\eea
That the anomalous dimension of the scalar field makes an appearance should come
as no surprise: the WFGTI identity is at work. What is rather interesting about
the form (24) is that for $q^2=0$ it takes the form $[F(p^2)-F(k^2)]/(p^2-k^2)$
even though it is associated with the transverse piece; but for $q^2\neq 0$ this
particular structure disappears as one can see from the form of ${\cal D}$. Note
anyway that (23) and (24) both have intrinsic dependences on all three variables
$p^2, k^2$ and $q^2$ (or $p\cdot k$).

A third way of going non-perturbative relies upon dispersion relations; while these
are well-established for the two-point function, in the form of the 
Lehmann-K\"{a}ll\'{e}n representation, they are trickier for the vertex function
but can nevertheless be found as follows for graphs with triangular topology.
Make the change of variable $\sigma \rightarrow m^2/W^2$ in eq. (23) --- so that
$W^2$ runs from $m^2$ to $\infty$ --- in the denominator ${\cal D}$. This means
we can generally write
\bea
 S^{-1}(p) &=& \int_{m^2}^{\infty} dW^2\,\frac{\rho(W^2)}{[p^2-W^2+i\epsilon]},\nn\\
 \tau(p,k,q) &=& \int_{m^2}^{\infty} dW^2\int_{-1}^1 du\,
\frac{{\cal P}(W^2,u)}{[p^2(1+u)/2+k^2(1-u)/2+q^2(1-u^2)(W^2-m^2)/4-W^2+i\epsilon]}.
\label{third}
\eea
The idea is then to determine $\rho$ and ${\cal P}$ self-consistently through the
Schwinger-Dyson equations for the propagator and the vertex; the latter inevitably
brings in the 4-point function, but we can use its own WGFTI to approximate it
by connected 3-point graphs. While we have not solved this problem for 
${\cal P}$, the idea has been taken to fruition \cite{CNP} for $\rho$ in SQED and 
QED, giving results that coincide with perturbation theory up to order $e^4$ for
the charged field propagators. There is much more work involved in obtaining
the spectral functions ${\cal P}$ accurately to order $e^4$ and higher and this 
has not yet been done.
  These ans\"{a}tze are unlikely to be the whole story. However, one may ask 
how close these are to the real vertex and how these can be compared to each 
other. By construction, our ans\"atze agree with perturbtion at the one loop
order. 

\begin{itemize}

	\item {Ans\"atz}~(\ref{taudmnp}) agrees with perturbation theory to ${\cal O}(e^2)$ 
in all momentum regimes, dimensions and gauges. To see whether this relation
between the vertex and the propagator survives at ${\cal O}(e^4)$, one needs
to know these Green functions at that order.

\item {Ans\"atz}~(\ref{Anom-vert}) would be in accord with the real vertex in
4 dimensions for large
$k^2$ and $p^2$ but would fail for $k^2$ and $p^2 \approx m^2$. As one
knows $\gamma$ to ${\cal O}(e^2)$, it amounts to knowing the vertex to all orders
in the leading logarithm approximation for asymptotically large values of momenta.

\item  {Ans\"atz}~(\ref{third}) would agree with perturbation theory
order by order depending upon the exact knowledge of the $\rho$ and
${\cal P}$ functions to a given order. In principle, one could evaluate 
these functions non-perturbatively through SDEs. However, this exercise for
${\cal P}$ is a hard nut to crack.

\end{itemize}

A full 2-loop calculation of 
the vertex should narrow down possible forms of any ansatz. Techniques for the 
2-loop vertex calculation, especially for the massless particles, have been 
developed  in~\cite{Tarasov, Davyetal, Remiddi}. All the master integrals for 
massless 2-loop vertex diagram with three off-shell legs have been calculated 
in~\cite{Nigel}. These advances indicate that the calculation of the two-loop
transverse vertex should not be too difficult, at least for the massless case. 
This work is under progress.

\section{On The Transverse Takahashi Identities}

Eq.~(\ref{taudmnp}) is effectively a Ward-identity type relation linking the
transverse vertex to the scalar propagator. There have been attempts to look
for formal relations of this kind. Takahashi~\cite{Takahashi} discovered 
what are called transverse identities whose implications for the vertex have 
been examined for spinor QED~\cite{TWI1, TWI2, TWI3, TWI4}.
In case of SQED, as there is just one unknown which remains undetermined by the 
conventional WFGTI, it is tempting to look for a transverse Takahashi identity, 
hoping one might be able to determine the three-point vertex more realistically.

It should be noted that the general form of the vertex (7) shows that the 
transverse coefficient contributes to both basis vectors $P_\mu=(p+k)_\mu$ and 
$q_\mu=(k-p)_\mu$. The curl of the vertex $q_\mu\Gamma_\nu-q_\nu\Gamma_\mu$ will
eliminate the component of $\Gamma$ proportional to $q$, leaving us with the
kinematic mixture $q_\mu P_\nu - q_\nu P_\mu$, multipling the coefficient
\[
\frac{S^{-1)}(p)-S^{-1}(k)}{p^2-k^2} + q^2\tau(p^2,k^2,q^2)/2,
\]
which is unwieldy. However the same kinematic combination can also be obtained 
by forming the `modified curl' $P_\mu\Gamma_\nu - P_\nu\Gamma_\mu$; this has 
the advantage of killing off the longitudinal part of $\Gamma$ and {\em only 
bringing in} $\tau$. We suggest that new identities involving the modified curl 
are more appropriate and will prove more promising for SQED.

\section*{ACKNOWLEDGMENTS}

We thank A. Raya and M.E. Tejeda for valuable discussions. We acknowledge CIC 
and CONACyT for their financial support under grants number 4.10 and 46614-I.

\section*{APPENDIX}

In this appendix we summarise the results for various integrals
involved in the calculation of the 3-point vertex for quick
reference. We write down the results for arbitrary $d$ as well as
small $\epsilon=2-\ell$ and $d=3$. This way we aim to present all 
the integrals including the ones which have not been considered in the
articles~\cite{BC,KRP,BKP1,BKP2,BR1}. Wherever possible, we compare
the results of specific cases with the ones in the above-mentioned
articles.  \\

\noindent
{\bf{Subsidiary quantities~:}} \\

\noindent
There are a number of quantities which arise in various integrations
that constantly appear later on, so we shall summarise them first and invoke 
them as they turn up. Unless specified we work in general dimension 2$\ell$~:
\[
{\cal S}=\sqrt{1-\frac{4m^2}{q^2}}\ln\Bigg(\frac{\sqrt{1-
 \frac{4m^2}{q^2}}+1}{\sqrt{1-\frac{4m^{2}}{q^{2}}} - 1}\Bigg)
=2\sqrt{4m^2/q^2-1}\,\,\arctan\left(1/\sqrt{4m^2/q^2-1}\right)
\]
\[
I(p^2) \equiv (1/\sqrt{-p^2})\,\arctan\sqrt{-p^2/m^2},
\]
\[
L(p^2) \equiv (1-m^2/p^2)\,\ln(1-p^2/m^2),
\]
\[
Q_1(k) \equiv (\pi m^2)^{\ell-2}\Gamma(1-\ell)\,{}_2F_1(2-\ell,1;\ell;k^2/m^2),
\]
\[
\ell Q_2(k)\equiv(\pi m^2)^{\ell-2}\Gamma(2-\ell)\,{}_2F_1(2-\ell,1;\ell+1,k^2/m^2),
\]
\[
m^2Q_3(k)\equiv(\pi m^2)^{\ell -2}\Gamma(1-\ell)\,{}_2F_1(3-\ell,2;\ell,k^2/m^2),
\]
\[
m^2Q_4(k)\equiv(\pi m^2)^{\ell-2}(2-\ell)\Gamma(-\ell)\,{}_2F_1(3-\ell,2;\ell+1;k^2/m^2),
\]
\[
Q_5(k)\equiv-2i\pi^2(\pi m^2)^{\ell-2}\Gamma(2-\ell)\,{}_2F_1(1/2,1;\ell-1;4m^2/q^2),
\]
\[
Q_6(p)\equiv i\pi^\ell(\ell-3)(m^2)^{\ell-3}\Gamma(1-\ell)\,{}_2F_1(1,4-\ell;\ell;p^2/m^2)
\]
\[
\chi\equiv m^2(k^2-p^2)^2 + (m^2-k^2)(m^2-p^2)q^2.
\]
$\chi$ represents $(12\times$volume)$^2$ of a tetrahedron constructed with base 
triangle lengths $\sqrt{-k^2},\sqrt{-p^2},\sqrt{-q^2}$ and lateral sides to the 
apex of lengths $m,m,0$; thus it has geometrical significance. It is worthwhile 
noting the zero mass limits of the $Q_i$ as we will consider such situations later~:
\[
Q_1^{d,0}(k)=-(-\pi k^2)^{\ell-2}\Gamma^2(\ell-1)\Gamma(2-\ell)/\Gamma(2\ell-2)
 = -Q_2^{d,0}(k),
\]
\[
2\pi Q_3^{d,0}(k)=(-\pi k^2)^{\ell-3}\Gamma^2(\ell-2)\Gamma(3-\ell)/\Gamma(2\ell-4)
 = \pi Q_4^{d,0}(k)
\]
\[
(m^2)^{2-\ell}Q_5^{d,0}(k) \rightarrow -2i\pi^\ell\Gamma(2-\ell),
\quad Q_6^{d,0}(k)/m^2 \rightarrow i\pi^\ell(-k^2)^{\ell-4}
  \Gamma(\ell)\Gamma(\ell-2)/\Gamma(2\ell-4)
\]
\[
\chi^{d,0} = k^2p^2q^2 {\rm ~very~simply}.
\]
\\

\noindent
{\bf{The $K$ Integrals~:}} \\

\noindent $K^{(0)}$ in the list (\ref{TheIntegrals}) for arbitrary 
$d = 2\ell$ equals
\bea i\pi^2K_0/2 = K^{(0)} &=& i\pi^{\ell} \; \Gamma \left(2 - \ell\right) \; 
(m^{2})^{\ell -2} \;{}_2F_{1}\bigg(1, \; 2-\ell \; ; \; \frac{3}{2} \; ; \; 
 \frac{q^{2}}{4m^{2}} \; \bigg) \;,  \\ \nn \\
K^{(0)}_{d,0} &=& i\pi^{\ell}(-q^{2})^{\ell - 2}\frac{\Gamma^{2}\big(\ell
  - 1\big)\Gamma\big(2 -\ell \big)}{\Gamma\big(2\ell - 2\big)} \;.
\eea
Corresponding massive and massless expressions in the neighbourhood of 4-d
($\epsilon= 2-\ell$) are~:
\bea
K^{(0)}_{4-2 \epsilon,m} = i\pi^{2}[C - {\cal S}]\;,  \qquad \qquad
K^{(0)}_{4-2 \epsilon,0} = i\pi^{2}[C -\ln(-q^2/m^2)] \;,  \label{K04}
\eea
where
\begin{eqnarray}
 C=\frac{1}{\epsilon}-\gamma-\ln (\pi m^{2})+2\;, \qquad \;.
\label{CandS}
\end{eqnarray}
The first of the results (\ref{K04}) agrees with Eqs.~(44, 46-48) of~\cite{KRP}.
When $d=3$, this integral simplifies even more~:
\bea
 K^{(0)}_{3,m} = {i\pi^{2}} I(q^2/4),\;\qquad
 K^{(0)}_{3,0} = {i\pi^{3}}/{\sqrt{-q^{2}}}\;\qquad .\label{K03}
\eea
Expressions (\ref{K03}) coincide with (A1) of~\cite{BKP2} and (A3) of ~\cite{BR1} 
respectively. For the $K_\mu^{(1)}$ integral in (12), it is easy to show that 
$K_{\mu}^{(1)} = (p + k)_\mu \; K^{(0)}/2$. \\

\noindent
{\bf{The $J$ Integrals~:}} \\

\noindent The $J^{(0)}$ integral can be found in various 
sources~\cite{KRP,BKP0,BKP2,BR1,AID}
but the most general case (massive mesons and any $d$) is not known; we shall 
simply cite the known answers. $J_0$ is probably the most difficult one to 
work out as it brings in dilogarithmic or Spence ($Sp$) functions when $\ell$ 
is integer. For any $\ell$ the massless case has been given a completely elegant
representation by Davydychev~\cite{AID} :
\bea
J^{(0)}&=&2i(-i\pi)^\ell(k^2,p^2,q^2)^{\ell-2}\frac{\Gamma^2(\ell-1)\Gamma(2-\ell)}
   {\Gamma(2\ell-2)}\bigg[\frac{(p^2q^2)^{2-\ell}}{p^2+q^2-k^2}\,
   {}_2F_1\bigg(1,1/2;\ell-1/2,-\frac{\Delta}{2(p^2+q^2-k^2)^2}\bigg)\nn\\
 &&\hspace{2.5in} + {\rm two~perms} - \pi\frac{\Gamma(2\ell-2)}{\Gamma^2(\ell-1)}
 (2\Delta)^{\ell-3/2}\Theta \bigg]   \label{J00}
\eea
For massive mesons, like we have, the result $J_0^{4,m}$ in 4-D is too lengthy (and
uninformative) to quote. It is given in eq. ((16) of ref.\cite{KRP} and involves
Spence functions of complicated arguments. In 3-D the result is~\cite{BR1} easy to 
state~:
\bea 
J_0^{3,m} &=& \eta(k,p)I(\eta^2(k,p)\chi/4)+\eta(p,k)I(\eta^2(p,k)\chi/4); \quad
\eta(k,p) = \frac{m^2(k^2-p^2)(2m^2-k^2-p^2)+\chi}{\chi(m^2-k^2)}.
\eea
In the massless limit one obtains~\cite{BKP0,BKP2} from all of these forms,
\bea
J_0^{4,0} &=& \frac{2}{\Delta}\bigg[Sp \bigg(\frac{p\cdot q-\Delta}{-p^2}\bigg) -
 Sp\bigg( \frac{p\cdot q+\Delta}{-p^2} \bigg) + \frac{1}{2}
 \bigg( \frac{p^2+p\cdot q -\Delta}{p^2+p\cdot q+\Delta}\bigg)
 \ln\bigg(\frac{q^2}{p^2}\bigg)\bigg],
\eea
\bea
J_0^{3,0} &=& \pi({\cal K}^2 + {\cal P}^2)/{\cal K}{\cal P}{\cal Q};\quad
 {\cal K}\equiv\sqrt{-k^2}, {\cal P}\equiv\sqrt{-p^2}, {\cal Q}\equiv\sqrt{-q^2} 
\eea

\noindent {\bf The $J_{\mu}$ Integral~:}\\

\noindent In its most general form, this can be written as 
\bea 
J_{\mu}^{(1)} &=& \frac{i\pi^{2}}{2}[ \; k_{\mu}
\; J_{A}(k,p) + p_{\mu} \; J_{B}(k,p) \; ] \; ;\qquad
 J_A(k,p)=J_B(p,k). \nn 
\eea 
We find 
\bea
J_{A}(k,p)&=&-\frac{1}{2\Delta^{2}}\bigg\{  \left[p^{2} - k \cdot p\right]K_{0} 
+ \left[p^{2}(k^{2} - m^{2}) - k \cdot p(p^{2} - m^{2}) \right]J_{0}+ 
2p^2Q_{1}(p) - 2k \cdot p\, Q_{1}(k) \bigg\} \;. \label{Jmu1dm} 
\eea 
In the massless case, 
\bea
J_{A}^{d,0}(k,p)=-\frac{1}{2\Delta^{2}}\bigg\{[p^2 -k\cdot p]K^{d,0}_0
 + p^2[k^{2}- k \cdot p]J_{0}^{d,0} + 2p^{2}Q_{1}^{d,0}(p) 
-2k\cdot p Q_{1}^{d,0}(k) \bigg\}\; .
\eea 
For small $\epsilon=2-\ell$, Eq.~(\ref{Jmu1dm}) reduces to the following 
expressions (see (A16, A17, 3.9(a,b,c)) of~\cite{BC} and (39 - 44) of~\cite{KRP}) 
\bea 
J_A^{4-2\epsilon,m}&=&\frac{1}{2\Delta^2}\bigg\{-(m^2 p\cdot q + p^2 k\cdot q)
J_0^{4-2\epsilon,m} + 2k\cdot p L(k)-2p^2 L(p)-2p\cdot q {\cal S} \bigg \}\;
\nn \\
J_A^{4-2\epsilon,0}&=&\frac{1}{2\Delta^2}\bigg\{ p^2(k\cdot p-k^2)
J_0^{4 -2\epsilon,0} +2(p^2 -k\cdot p)\ln(-q^{2})-2p^2\ln(-p^2)+
2k\cdot p\ln(-k^2)\bigg\}\;,
\eea 
Similar results for $d=3$ are~:
\bea 
J_{A}^{3,m}&=&\frac{1}{2\Delta^{2}}\bigg\{[p^{2}(k\cdot p-k^2) +
m^{2}(p^{2} - k\cdot p)]J_{0}^{3,m} + 2(k\cdot p-p^2)I(q^2/4)
+ 4p^2  I(p^2) -4k\cdot p I(k^2) \bigg\}\; ,   \nn \\ \nn \\
J_{A}^{3,0}&=& \frac{1}{2\Delta^{2}}\bigg\{[p^{2}(k\cdot p-k^2)]
J_{0}^{3,0} -\frac{2\pi}{\sqrt{-q^{2}}}(p^{2} -k\cdot p)
+\frac{2\pi p^2}{\sqrt{-p^2}} - \frac{2\pi k\cdot p}{\sqrt{-k^2}}\bigg\}\,.
\eea
The first of these expressions coincides with Eq.~(A5) of~\cite{BR1} and the 
second with Eq.~(A3) of~\cite{BKP2} after appropriate change of notation. \\

\noindent
{\bf{The $J_{\mu \nu}^{(2)}$ Integral~:}} \\

\noindent
From symmetry considerations, the integral $J_{\mu \nu}^{(2)}$ of the list 
(\ref{TheIntegrals}) can be expanded out as follows~:
\bea
J_{\mu\nu}^{(2)}
&=& \frac{i\pi^{2}}{2}\bigg[\frac{g_{\mu\nu}}{2\ell}K_0 + \bigg(k_{\mu}k_{\nu} 
- g_{\mu\nu}\frac{k^{2}}{2\ell}\bigg)J_C + 
\bigg(p_\mu k_\nu + k_\mu p_\nu - g_{\mu\nu}\frac{(k \cdot p)}{\ell}\bigg)J_{D} 
+ \bigg(p_\mu p_\nu -g_{\mu\nu}\frac{p^{2}}{2\ell}\bigg)J_{E}\bigg]\;. \nn
\eea
The coefficients $J_C,\,J_D$ and $J_E$ in the above expressions are~: 
\bea
J_{C}(k,p)&=&\frac{1}{4(\ell - 1)\Delta^2}\bigg\{
 [(2\ell-2)(p^2-m^2)k\cdot p - (2\ell-1)(k^2 -m^2)p^2] J_A -\; \nn \\
 &&\hspace{2cm} (p^{2} -m^{2})p^{2}J_{B} + 
 [(2-\ell)p^{2} + (\ell-1) k\cdot p]K_{0}-4(\ell-1)k\cdot pQ_2(k)\bigg\}\;,\nn\\ \nn \\
J_{E}(k,p) &=& J_C(p,k) \;,  \nn \\ \nn 
J_D(k,p)   &=& \frac{1}{8(\ell- 1)\Delta^{2}}
\bigg\{[2\ell k\cdot p (k^{2} - m^{2}) +(2 - 2\ell)k^2(p^2- m^2)]J_{A}  +
[2\ell (p^2 -m^2)k\cdot p + (2 - 2\ell)p^2(k^2 - m^2)]J_{B} \nn \\
&& \hspace{2cm} -[(\ell-1)q^2 +2k\cdot p]K_0 + 
4(\ell -1)[k^{2}Q_{2}(k) + p^{2}Q_{2}(p)]\bigg\}\;. \label{JCJDJE} 
\eea 
In the massless case we obtain the following expressions: 
\bea 
J_{C}^{d,0}(k,p)&=&\frac{1}{4(\ell- 1)\Delta^{2}}\bigg\{
[(2\ell-2)p^2k\cdot p-(2\ell-1)k^{2}p^{2}\bigg]J_{A}^{d,0}\nn \\
 && \hspace{2cm}- p^{4}J_{B}^{d,0} + [(2 -\ell)p^2+(\ell - 1)k\cdot p]K_{0}^{d,0} 
-4(\ell - 1)k\cdot pQ_{2}^{d,0}(k)\bigg\} \;, \nn \\ \nn \\ \nn
J_{D}^{d,0}(k,p)&=&\frac{1}{8(\ell - 1)\Delta^{2}}\bigg\{[2\ell k\cdot p k^2 
+ (2 - 2\ell)k^{2}p^{2}]J_{A}^{d,0} +
 [2\ell p^{2}k\cdot p + (2 - 2\ell)p^{2}k^{2}]J_{B}^{d,0}\nn\\
&& \hspace{2cm} - [(\ell - 1)q^{2} +2k\cdot p]K_{0}^{d,0} +
 4(\ell - 1)[k^{2}Q_{2}^{d,0}(k) + p^{2}Q_{2}^{d,0}(p)]\bigg\}\;. 
\eea 
Then for small $\epsilon=2-\ell$, we arrive at (compare
Eq.~(\ref{JCJD4m}) with Eqs.~(A18-A20) of~\cite{BC} and Eqs.~(49-51)
of~\cite{KRP}) 
\bea
J_{C}^{4-2\epsilon,m}&=&\frac{1}{4\Delta^{2}}\bigg\{2p^{2} +
2m^2 k\cdot p/k^2 -2k\cdot p{\cal S} +
2(k\cdot p)(1 - m^{2}/k^{2})L(k) +  \nn \\
&& \hspace{1cm} [2k\cdot p(p^{2} -m^{2})+ 3(m^{2} -k^{2})p^{2}]J_{A}^{4-2\epsilon,m} + 
p^{2}(m^{2} -p^{2})J_{B}^{4-2\epsilon,m}\bigg\}\; \nn \\  \nn \\
J_{D}^{4-2\epsilon,m}&=&\frac{1}{4\Delta^{2}}\bigg\{2k\cdot p
[(k^2-m^2)J_{A}^{4-2\epsilon,m} + (p^2-m^2)J_{B}^{4-2\epsilon,m}- 1] -
[m^2-k^2{\cal S} + (k^2-m^2)L(k)\nn\\
&& \hspace{1cm} + k^2(p^{2} - m^{2})J_{A}^{4-2\epsilon,m}] -
[m^2-p^2{\cal S}+(p^2-m^2)L(p)+ p^2(k^2-m^2)J_B^{4-2\epsilon,m}\bigg]\bigg\} \;. 
\label{JCJD4m}  \\ \nn\\
J_{C}^{4-2\epsilon,0}&=&\frac{1}{4\Delta^{2}}\bigg\{p^2[2k\cdot p-3k^2]
J_{A}^{4-2\epsilon,0}-p^{4}J_{B}^{4-2\epsilon,0} +2p^{2} + 
2k\cdot p\ln(k^{2}/q^{2})\bigg\} \nn \\ \nn \\
J_{D}^{4-2\epsilon,0}&=&\frac{1}{4\Delta^{2}}\bigg\{
[k^2(2k\cdot p-p^{2}]J_{A}^{4-2\epsilon,0}+[p^2(2k\cdot p-k^{2}]J_B^{4-2\epsilon,0} 
+ k^{2}\ln(q^{2}/k^{2})+p^{2}\ln(q^{2}/p^{2})  -2k\cdot p\bigg\}. 
\eea
Also for $d=3$,
\bea
J_{C}^{3,m}&=&\frac{1}{2\Delta^{2}}\bigg\{[(p^2-m^2)k\cdot p-2p^2(k^2-m^2)]
J_A^{3,m} -p^{2}(p^2-m^2)J_B^{3,m} + \frac{2k\cdot p} {k^2}(m^2 -k^2)I(k^2) 
+(k\cdot p + p^{2}) I(q^2/4)
\nn \\
&& \hspace{1cm}
- 2m\frac{k\cdot p}{k^{2}}\bigg\}  \nn \\ \nn \\
J_{D}^{3,m}&=&\frac{1}{4\Delta^{2}}\bigg\{[k^{2}(3k\cdot p -p^{2})
-m^2(3k\cdot p -k^2)]J_A^{3,m}+[p^2(3k\cdot p-p^2)-m^2(3k\cdot p-p^2)]J_B^{3,m}
 \nn \\ 
&& \hspace{1cm} -2(m^2 -k^2) I(k^2) - 2(m^2 -p^2)I(p^2)-(k+p)^2 I(q^2/4)+4m
  \bigg\}  \label{JCJD3m}  \\ \nn \\
J_{C}^{3,0}&=&\frac{1}{2\Delta^{2}}\bigg\{p^{2}(k\cdot p -2k^{2})J_{A}^{3,0} 
-p^{4}J_{B}^{3,0} - \pi k\cdot p/\sqrt{-k^{2}}+\pi p\cdot(p+k)/{\sqrt{-q^{2}}} 
\bigg\}   \nn \\ \nn \\
J_{D}^{3,0}&=&\frac{1}{4\Delta^{2}}\bigg\{k^{2}(3k\cdot p-p^{2})J_{A}^{3,0} 
+ p^{2}(3k\cdot p -p^{2})J_{B}^{3,0} +\pi k^{2}/\sqrt{-k^2}+\pi p^2/\sqrt{-p^2} 
- \pi(k + p)^{2}/\sqrt{-q^2}   \bigg\} \;. \label{JCJD30} 
\eea
Eq.~(\ref{JCJD3m}) is in agreement with (A6, A7) of~\cite{BR1} and Eq.~(\ref{JCJD30}) with (A4) of~\cite{BKP2}. \\

\noindent
{\bf{The $I^{(0)}$ Integral~:}} \\

\noindent
The massive $I_{0}$ integral in arbitrary dimensions is given by
\bea
I^{(0)} &=&\frac{1}{2 \chi}\bigg \{4(2-\ell)q^{2}(m^{2}+ k \cdot p )J^{(0)} 
+ (4 m^{2} - q^{2})Q_5(q) \nn
\\ && \hspace{0.95cm}  + [(q^2- 2m^2)(p^2- m^2) + 2m^2(k^2 -m^2)]Q_6(p)\nn \\ 
&& \hspace{0.95cm}  + [(q^2- 2m^2)(k^2-m^2) + 2m^2(p^2 - m^2)]Q_6(k)\bigg\} \;
\eea
Hence the massless case reduces to 
\bea
I^{(0)}_{d,0}=\frac{i \pi^2}{k^{2}p^{2}}\bigg[(2 - \ell)k\cdot pJ_{0}^{d,0} 
- k^{2}Q_{3}^{d,0}(k) -
p^{2}Q_{3}^{d,0}(p)+ q^{2}Q_{3}^{d,0}(q)\bigg]\;. 
\eea 
When $\epsilon=2-\ell$ is small, we have 
\bea I^{(0)}_{4-2
\epsilon,m}&=&i\pi^{2}\bigg \{\frac{1}{\chi}\bigg[- q^{2}{\cal S} +
p^{2}\frac{[(p^{2} - m^{2})q^{2} + 2 m^{2}(k^{2} - p^{2})]}{(p^{2} -
m^{2})^{2}}L(p) + k^{2}\frac{[(k^{2} - m^{2})q^{2} - 2 m^{2}(k^{2} -
p^{2})]}{(k^{2} - m^{2})^{2}}L(k) \bigg] \nn \\ && \hspace{1.5cm}
- \frac{C-2}{(p^{2} - m^{2})(k^{2} - m^{2})}\bigg \} \; \nn \\ \nn \\
I^{(0)}_{4-2 \epsilon,0}&=&\frac{i\pi^{2}}{k^{2}p^{2}}
\bigg[2-C + \ln\bigg(\frac{k^{2}p^{2}}{q^{2}m^2}\bigg)\bigg] \;, 
\eea 
whereas for $d=3$, 
\bea 
I^{(0)}_{3,m}&=&\frac{1}{\chi}\bigg \{q^{2}(m^{2} +
k\cdot p)J^{(0)}_{3,m} + i\pi^{2}m\bigg[\frac{q^{2}(k^{2} - m^{2}) -
(k^{2} - p^{2})(k^{2} + m^{2})}{(k^{2} - m^{2})^{2}}
+ \frac{q^{2}(p^{2} - m^{2}) + (k^{2} - p^{2})(p^{2} +
m^{2})}{(p^{2} - m^{2})^{2}}\bigg] \bigg \}\; \nn \\ \nn \\
I^{(0)}_{3,0}&=&\frac{k \cdot p}{k^{2}p^{2}}J^{(0)}_{3,m}\;.
\label{I03} 
\eea
The first of Eq.~(\ref{I03}) agrees with (A8) of~\cite{BR1}. \\

\noindent
{\bf{The $I_{\mu}^{(1)}$ Integral:}} \\

\noindent
In analogy with the integral $J_{\mu}^{(1)}$, $I_{\mu}^{(1)}$ can be
expanded out as 
\bea 
I_{\mu}^{(1)}&=&\frac{i
\pi^{2}}{2}[k_{\mu}I_{A}(k,p) + p_{\mu}I_{B}(k,p)];\qquad I_B(k,p) = I_A(p,k),
\eea 
where
\bea 
I_{A}(k,p)&=&\frac{1}{2\Delta^{2}}\bigg\{[k\cdot p(p^2-m^2)-p^2(k^2-m^2)]I_0 
+ [k\cdot p -p^2]J_0 + 2 k\cdot pQ_{3}(k) - 2p^2Q_3(p) \bigg\}\;. \nn
\eea
In the massless case  
\bea
I_{A}^{d,0}(k,p)=\frac{1}{2\Delta^{2}}\bigg\{p^{2}[k\cdot p-k^2]I_0^{d,m} 
+ [k \cdot p - p^2]J_{0}^{d,m}+2k\cdot pQ_3^{d,0}(k) -2p^2Q_3^{d,0}(p)\bigg\}\;. 
\eea
Near 4-dimensions
\bea 
I_{A}^{4-2\epsilon,m}&=&\frac{1}{2\Delta^{2}}\bigg\{-k\cdot qJ_0^{4-2\epsilon,m} 
- 2q^{2}(m^{2} -k^{2})(k^{2} - k\cdot p){\cal S}/\chi \nn \\ 
&& \hspace{1cm} + \frac{2L(p)}{(m^{2} - p^{2})}[p^{2} - k\cdot p+p^{2}q^{2}
(k^2-m^2)(m^2+k\cdot p)/\chi]+2k^2q^2(m^2+k\cdot p)L(k)/\chi\bigg\}\;,\label{IA4m}\\
I_{A}^{4-2 \epsilon,0} &=& \frac{1}{2\Delta^{2}}\bigg\{ [k\cdot p -p^2]
 J_{0}^{4-2 \epsilon,0} -2[k \cdot p - k^{2}]\frac{\ln(-q^{2})}{k^{2}} 
 + 2\frac{k\cdot p}{k^{2}}\ln (-p^{2})- 2\ln(-k^{2}) \bigg \} \;. 
\eea 
Eq.~(\ref{IA4m}) is in agreement with the expression (53) of~\cite{KRP}. 
Similar answers for $d=3$ are 
\bea 
I_{A}^{3,m}&=&\frac{2}{\Delta^{2}}\bigg\{[k\cdot p(p^{2} - m^{2}) - 
p^{2}(k^{2} - m^{2})]\frac{I_{0}^{3,m}}{4}
+ [k \cdot p - p^{2}]\frac{J_{0}^{3,m}}{4}
  + \frac{mp^{2}}{(m^{2} - p^{2})^{2}} - \frac{m k \cdot p}
  {(m^{2} - k^{2})^{2}} \bigg\}\;, \label{IA3m}   \\
I_{A}^{3,0}&=&-\frac{\pi}{k^{2}\sqrt{-q^{2}k^{2}p^{2}}} \label{IA30}  \;.
\eea
Eq.~(\ref{IA3m}) agrees with (A11) of~\cite{BR1}. \\

\noindent
{\bf{The $I_{\mu \nu}^{(2)}$ Integral:}} \\

\noindent
The integral $I_{\mu \nu}^{(2)}$ of the list (\ref{TheIntegrals}) 
may be decomposed as follows~:
\bea
I^{(2)}_{\mu \nu}
&=&\frac{i\pi^{2}}{2}\bigg[\frac{g_{\mu\nu}}{2\ell}J_{0} +
\bigg(k_{\mu}k_{\nu} - g_{\mu\nu}\frac{k^{2}}{2\ell}\bigg)I_{C} +
\bigg(p_{\mu}k_{\nu} + k_{\mu}p_{\nu} - g_{\mu\nu}\frac{k \cdot p}{\ell}
\bigg)I_{D}+\bigg(p_{\mu}p_{\nu}-g_{\mu\nu}\frac{p^{2}}{2\ell}\bigg)I_{E}\bigg]\;,
\\ I_E(k,p) &=& I_C(p,k),
\eea
where, in arbitrary dimensions,
\bea
I_{C}(k,p)&=&\frac{1}{4(\ell - 1)\Delta^{2}}\bigg\{[(2\ell -2)(p^2-m^2)k\cdot p
- (2\ell -1)(k^{2} - m^{2})p^{2}]I_{A} -(p^2- m^2)p^2I_{B}\nn \\ 
&& \hspace{2cm} +[(2\ell-2)k\cdot p-(2\ell -1)p^2]J_A
-p^2J_{B} + 2p^{2}J_0+4(\ell-1)k\cdot pQ_4(k)\bigg\}\;,
\\ \nn \\
I_{D}(k,p)&=&\frac{1}{8(\ell\!-\!1)\Delta^{2}}\bigg\{[2\ell (k^2-m^2)k\cdot p-
(2\ell-2)(p^2-m^2)k^2]I_A+[2\ell(p^2-m^2)k\cdot p-(2\ell\!-\!2)(k^2-m^2)p^{2}]I_B 
\nn \\
&&  \hspace{1.5cm} +[2\ell k\cdot p\!-\!(2\ell-2)k^2]J_A\!-\!4k\cdot pJ_0\!+\!
[2\ell k\cdot p\!-\!(2\ell\!-\!2)p^2]J_B\!-\!
4(\ell\!-\!1)[p^{2}Q_4(p)\!+\!k^{2}Q_4(k)]\bigg\}\;. 
\eea
For the massless case, we arrive at the following simplified results
\bea 
I_{C}^{d,0} &=&\frac{1}{4(\ell-1)\Delta^{2}} \bigg\{p^{2}
[(2\ell - 2) k\cdot p - (2\ell -1) k^2]I_{A}^{d,0}- p^4 I_B^{d,0} 
+ [(2\ell-2)k\cdot p-(2\ell -1)p^2]J_A^{d,0}- p^{2}J_{B}^{d,0}\nn\\
&& \hspace{2cm} +2p^2 J_{0}^{d,0}+4(\ell-1)k\cdot p\,Q_4^{d,0}(k)\bigg\}\;,\nn \\
I_{D}^{d,0}&=&\frac{1}{4(\ell - 1)\Delta^{2}} \bigg\{[\ell k\cdot p
- (\ell - 1)p^{2}](J_{B}^{d,0}+ k^{2}I_{A}^{d,0}) + 
 [\ell k\cdot p-(\ell-1)k^2](J_{A}^{d,0}+p^{2}I_{B}^{d,0}) \nn \\
&&\hspace{2cm}-2k\cdot pJ_0^{d,0}-2(\ell-1)[p^2Q_4^{d,0}(p)+k^2Q_4^{d,0}(k)]\bigg\}\;. 
\eea 
Near 4-dimensions, these expressions yield 
\bea 
I_{C}^{4-2 \epsilon,m}&=&\frac{1}{4\Delta^{2}}\bigg
\{2p^{2}J_{0}^{4-2 \epsilon,m} - \frac{4k \cdot p}{k^{2}}\bigg(1 +
\frac{m^{2}L(k)}{(k^{2} - m^{2})}\bigg) + \{2k\cdot p -
3p^{2}\}J_{A}^{4-2 \epsilon,m} - p^{2}J_{B}^{4-2 \epsilon,m} +  \nn \\
&& \hspace{1cm} [-2k\cdot p(m^2 - p^2)+ 3p^{2}(m^2-k^2)]I_{A}^{4-2\epsilon,m}+
p^{2}(m^{2} - p^{2})I_{B}^{4-2 \epsilon,m} \bigg \}\;,  \nn \\
I_{D}^{4-2 \epsilon,m}&=&\frac{1}{4\Delta^{2}}\bigg \{ -2(k \cdot
p)J_{0}^{4-2 \epsilon,m} + 2\bigg(1 + \frac{m^{2}L(k)}{(k^{2} -
m^{2})}\bigg ) + 2\bigg(1 + \frac{m^{2}L(p)}{(k^{2} - m^{2})} \bigg)+
(2k\cdot p - k^{2})J_{A}^{4-2 \epsilon,m} \nn \\ &&  \hspace{1cm}+
(2k\cdot p - p^{2})J_{B}^{4-2 \epsilon,m} + [k^{2}(m^{2} - p^{2}) -
2k \cdot p(m^{2} - k^{2})]I_{A}^{4-2 \epsilon,m}
 \nn\\&&\hspace{1cm}+[p^2(m^2-k^2)-2k\cdot p(m^2-p^2)]I_B^{4-2\epsilon,m}\bigg\} \;,  \\
I_{C}^{4-2 \epsilon,0}&=&\frac{1}{4\Delta^{2}}\bigg
\{2p^{2}J_{0}^{4-2 \epsilon,0} - 4\frac{k \cdot p}{k^{2}} + (2k\cdot
p - 3p^{2})J_{A}^{4-2 \epsilon,0} - p^{2}J_{B}^{4-2 \epsilon,0}  -
p^{4}I_{B}^{4-2\epsilon,0}+ p^2(2k\cdot p-3k^{2})I_{A}^{4-2\epsilon,0} \bigg \}\;,  \nn \\
I_{D}^{4-2\epsilon,0}&=&\frac{1}{4\Delta^{2}}\bigg \{ -2k \cdot
pJ_{0}^{4-2 \epsilon,0} + 4 + (2k\cdot p - k^{2})J_{A}^{4-2
\epsilon,0} + (2k\cdot p - p^{2})J_{B}^{4-2 \epsilon,0}
- k^{2}( p^{2} - 2k \cdot p)I_{A}^{4-2 \epsilon,0} \nn \\ 
&& \hspace{1cm} -p^{2}(k^{2} - 2k\cdot p)I_{B}^{4-2 \epsilon,0} \bigg
\}\;. 
\eea 
Finally for $d=3$, we have 
\bea
I_{C}^{3,m}&=&\frac{1}{2\Delta^{2}}\bigg\{2p^{2}J_0^{3,m} +
[p^2(k\cdot p-2k^2)-m^2(k\cdot p-2p^2)]I_{A}^{3,m} -
p^{2}(p^{2}-m^{2})I_{B}^{3,m} + (k\cdot p-2p^2)J_{A}^{3,m}-p^{2}J_B^{3,m} \nn\\ 
&& \hspace{1cm} + \frac{2m k\cdot p}{k^{2}(m^2-k^2)}-\frac{2k\cdot p}{k^2}
I(k^2)  \bigg\} \;, \nn \\
I_{D}^{3,m}&=& \frac{1}{4\Delta^{2}}\bigg\{ -4k\cdot pJ_0^{3,m}+ 
[k^{2}(3k\cdot p - p^2) - m^2(3 k\cdot p-k^2)]I_{A}^{3,m}
+[p^2 (3k\cdot p - k^2) - m^2(3 k\cdot p-p^2)]I_{B}^{3,m} \nn \\ 
&& \hspace{1cm} +(3k\cdot p-k^2)J_{A}^{3,m}+(3k\cdot p-p^2)J_{B}^{3,m}
- \frac{2m}{m^{2} - k^{2}} -\frac{2m}{m^2-p^2}+2I(k^2)+2I(p^2) \bigg\} \;,\nn\\
I_{C}^{3,0} &=& \frac{1}{2\Delta^{2}}\bigg\{p^{2}(k\cdot p -
2k^{2})I_{A}^{3,0}  -p^{4}I_{B}^{3,0} + (k\cdot p -
2p^{2})J_{A}^{3,0} - p^{2}J_{B}^{3,0}   - \frac{\pi k\cdot
p}{k^{2}\sqrt{-k^{2}}} -
\frac{4\pi p^{2}}{\sqrt{- k^{2}p^{2}q^{2}}} \bigg\} \;, \nn \\
I_{D}^{3,0}&=&\frac{1}{4\Delta^{2}}\bigg\{ k^{2}(3k\cdot p -
p^{2})I_{A}^{3,0} + p^{2}(3k\cdot p - k^{2})I_{B}^{3,0} +(3 k\cdot p
- k^{2})J_{A}^{3,0} + (3 k\cdot p - p^{2})J_{B}^{3,0} \nn \\ &&
\hspace{1cm} +\frac{\pi}{\sqrt{-k^{2}}} + \frac{\pi}{\sqrt{-p^{2}}}
+ \frac{8 k \cdot p}{\sqrt{- k^{2}p^{2}q^{2}}} \bigg\} \;. \eea

\noindent
{\bf{The $I_{\mu \alpha \beta}^{(3)}$ Integral:}} \\

\noindent
The integral $ I_{\mu \alpha \beta}^{(3)}$ comes contracted with vectors 
$p^\alpha$ and $k^\beta$ so it is straightforward to show that
\bea
-8 p^{\alpha}k^{\beta}I^{(3)}_{\mu \alpha \beta}
&=&  - 2 p^{\alpha}(k^{2} - m^{2})I^{(2)}_{\mu
\alpha} - 2 p^{\alpha}J^{(2)}_{\mu \alpha} +
i\pi^{2}p_{\mu}[Q_{2}(p) - (p^{2} - m^{2})Q_{4}(p)]
\nn \\ && \hspace{0cm}
- 2 k^{\beta}(p^{2} - m^{2})I_{\mu \beta}^{(2)} -
2k^{\beta}J_{\mu \beta}^{(2)} + i\pi^{2}k_{\mu}[Q_{2}(k)- (k^{2} -
m^{2})Q_{4}(k)]\;.
\eea
Therefore, we have $p^\alpha k^\beta I_{\mu\alpha\beta}^{(3)}$ in terms of
integrals we already know.

\end{document}